\documentclass{aa}

\def\dg{$^{\circ}$}

\def\logg{$\log g$}

\usepackage{float}
\usepackage{graphicx}

\usepackage[varg]{txfonts}
\usepackage[colorlinks = true,
            linkcolor = blue,
            urlcolor  = blue,
            citecolor = blue,
            anchorcolor = blue]{hyperref}

\begin{document}

\title{New insights on the massive interacting binary UU\,Cassiopeiae  }

\author{R.E. Mennickent\inst{1} \and G. Djura{\v s}evi{\'c}\inst{2} \and I. Vince\inst{2} \and  J. Garc\'es\inst{1} \and   P. Hadrava\inst{3} \and M. Cabezas\inst{3,4}  \and J. Petrovi\'c\inst{2} \and M. I. Jurkovic \inst{2,5} \and D. Kor\v{c}\'{a}kov\'{a} \inst{6}   \and H. Markov \inst{7} }

\institute{Universidad de Concepci\'on, Departamento de Astronom{\'{i}}a, Casilla 160-C, Concepci\'on, Chile \and 
Astronomical Observatory Belgrade, Volgina 7, 11060 Belgrade, Serbia  \and  
Astronomical Institute of the Academy of Sciences of the Czech Republic, Bo\v{c}n\'{\i} II 1401/1, 141 00 Praha 4, Czech Republic \and 
Institute of Theoretical Physics, Faculty of Mathematics and Physics, Charles University, V Hole\v{s}ovi\v{c}k\'{a}ch 2, 180 00 Praha 8, Czech Republic \and 
Konkoly Thege Astronomical Institute, Research Center for Astronomy and Earth Sciences, Konkoly Thege Mikl\'os \'ut 15-17, H-1121 Budapest, Hungary \and  
Astronomical Institute of Charles University, V Hole\v{s}ovi\v{c}k\'{a}ch 2, 180 00 Praha 8, Czech Republic  \and 
Institute of Astronomy and NAO, Rozhen, Smolyen, Bulgaria}

\date{Received 10 April 2020 / Accepted 1 May 2020}

 \titlerunning{On the massive interacting binary UU\,Cas}
\authorrunning{Mennickent et al.}

\abstract{We present the results of the study of the close binary UU Cassiopeiae based on previously published multi wavelength photometric and spectroscopic data. 
Based on eclipse timings of the last 117 years, we find an improved orbital period of $\rm P_{o} =  8\fd519296(8)$. In addition, we find a long cycle of length $T$ $\sim$ 270 d in the $I_c$-band data. 
There is no evidence for orbital period change during the last century, suggesting that the rate of mass loss from the system or mass exchange between the stars should be small. 
Sporadic and rapid brightness drops of up to $\Delta$$V$ = 0.3 mag are detected during the whole orbital cycle
and infrared photometry clearly suggests the presence of circumstellar matter.
We model the orbital light curve of 11 published datasets fixing the mass ratio and  cool star temperature 
from previous spectroscopic work; $q$=  0.52 and $T_c$= 22\,700\,K. 
We find a system seen at angle 74\dg with a stellar separation of 52 ${\rm R_{\odot}}$,
a temperature for the hotter star  $T_h$= 30\,200\,$K$  and stellar masses 17.4 and 9 ${\rm M_{\odot}}$ , radii 7.0 and 16.9 ${\rm R_{\odot}}$ 
and surface gravities \logg\ = 3.98 and 2.94,  for the hotter and cooler star, respectively. We find
an accretion disk surrounding the more massive star, with a radius of 21 ${\rm R_{\odot}}$ and vertical thickness in its outer edge of 6.5 ${\rm R_{\odot}}$, mostly
occulting the hotter star. Two active regions hotter than the surrounding disk are found, one located roughly in the expected position where
the stream impacts the disk and the other one in the opposite side of the disk. Changes are observed in parameters of the disk and spots in different datasets. 
}

\maketitle

\section{Introduction}

Massive close binaries are progenitors of many interesting objects as X-ray binaries and binary radio pulsars \citep{1993ARA&A..31...93V},  supernovae Ib and Ic \citep{1992ApJ...391..246P}, as well as collapsars resulting in gamma ray bursts \citep{2005A&A...435..247P} 
and gravitational wave sources \citep{2018MNRAS.481.1908K}. 
Massive binaries trace the evolutive history of these interesting objects
and can explain their galactic populations  \citep[e.g.][]{2020arXiv200207230Z}.
In addition, massive binaries are natural laboratories to understand the 
physical processes occurring during the periods of mass exchange and mass loss, that regulate
their evolution determining its final fate \citep{2014ApJ...782....7D}. 

The eclipsing close binary  UU\,Cassiopeiae  (UU\,Cas, BD +60 2629, 2MASS J23503951+6054391, 
$\alpha_{2000}$= 23:50:39.52, $\delta_{2000}$= +60:54:39.14, $V$ = 9.74, Spectral Type B0.5III)\footnote{http://simbad.u-strasbg.fr/simbad/}
has  a distance based on the GAIA DR2 parallax of 3256 [+392 -319] pc \citep{2018AJ....156...58B}.
It is listed in the catalogue of massive close binaries of \citet{2004A&AT...23..213P} that  compiles observable data for 176 massive close binaries with main sequence components earlier than approximately B5. In this catalogue, the object occupies the place 7th among 65
binaries with orbital period known, ordered in descending order of period length. 
UU\,Cas was studied by means of photographic spectra by \citet{1934ApJ....79...84S} who found an orbital period P$_{\rm o}$=  8\fd520676. 
The General Catalog of Variable Stars\footnote{http://www.sai.msu.su/gcvs/} \citep{2017ARep...61...80S} gives an orbital period of  8\fd51929 with reference to \citet{PK40}.

\citet{2002ARep...46..900P} obtained $UBVR$ differential magnitudes between 1984-1989 and
derived stellar masses of 34.5 $\pm$ 1.5 and 25.7 $\pm$ 0.6 $M_{\odot}$.
She also noticed  large magnitude deviations from the assumed over-contact binary model at some epochs 
which she has explained in terms of mass flows. In particular, she noticed the large 
variability of eclipse depths and overall light curve shape. The complex variability of the light curve and large deviations were ratified 
nine years later with new data  by \citet{2009arXiv0907.1047K}. Using 
photometric observations of UU\,Cas obtained until year 2000, the following ephemerides is given for the primary minimum by \citet{2004AcA....54..207K}\footnote{https://www.as.up.krakow.pl/o-c/index.php3}:
Min I = JD 2428751.6762 + 8\fd519281 E. 

\citet{2017AstBu..72..321G}  derived stellar masses of 17.7 and 9.5  $M_{\odot}$ assuming an orbital inclination of 69\dg and using medium resolution ($R$ = 15\,000) spectra. 
He also  presented evidence 
for a disk surrounding the hotter component and a common expanding envelope. 
Recent H$\alpha$ Doppler tomography revealed the importance of the gas flows in the semi-detached system to understand the  
different  flux contributions.  Gas stream from the cooler star, accretion disk around the hotter star and a wind
are taken into account by   \citet{2019ApJ...883..186K} to model the orbital variability of the H$\alpha$ profiles. These authors and  \citet{2017AstBu..72..321G}  noticed that the apparent paradox of a  deeper light minimum when
the more massive star is occulted by the less massive star can be explained if the more massive star is surrounded by an accretion disk. The presence of a disk 
was already suggested by  \citet{2010ASPC..435..301D} and 
\citet{2010POBeo..90..159M,2011BlgAJ..15...87M}.
The larger masses obtained by \citet{2002ARep...46..900P} compared with those of \citet{2017AstBu..72..321G} are  explained 
by the assumed over-contact system configuration and the absence of the disk in her model. 

In this study we hope to contribute to: (1) get for the first time a physical representation of the accretion disk  surrounding the more massive star
and at the same time get reliable and consistent stellar and system parameters, 
(2) shed light on the complex photometric variability of the system and 
the conflicting masses obtained from previous photometric and spectroscopic studies, (3)  use 
 the photometric datasets acquired over last years by all-sky surveys and satellites to advance in the understanding of this system, (4)
 improve the 
 value of the orbital period and check for its variability, 
and (5)  get insights on the system structure using published infrared photometric data.
   
The paper is organized as follows: in Section 2 we give details of the photometric observations analyzed in this paper, in Section 3 
we present our results based on photometric studies, in Section 4 we present the light curve model and their physical parameters, in Section 5 we discuss our findings in the context of earlier 
work on the system and in Section 6 we summarize our conclusions. 
 
\section{Observations}

\subsection{Photometric data and methodology}

The photometric data used in this paper were collected from previously published articles and publicly available databases. 
We have studied the photometric time series found in The All-Sky Automated Survey for Supernovae ASAS-SN\footnote{\url{http://www.astronomy.ohio-state.edu/asassn/index.shtml}} in $V$ and $g$,
the data from Kamogata Kiso Kyoto Wide-field Survey \citep{110009799479} in $I_{c}$ and $V$. In addition, we include
data acquired with The Optical Monitoring Camera \citep[OMC,][]{2003A&A...411L.261M} on board the high-energy INTEGRAL satellite, that provides photometry in the Johnson $V$-band within a 5 by 5 degree field of view
\citep{2010ASSP...14..493D} and it is able to detect optical sources brighter than around $V$ $\sim$ 18.
We also include  differential photometry from  
\citet[][$UBV$]{2009arXiv0907.1047K} and   \citet[][$UBVR$]{2002ARep...46..900P}.

We query for infrared magnitudes through the NASA/IPAC infrared science 
archive\footnote{\url{http://irsa.ipac.caltech.edu/Missions/wise.html}}.
We investigated the Wide-field Infrared Survey Explorer WISE, a NASA medium class explorer mission that conducted
an all-sky survey at mid-infrared bandpasses centered around wavelengths 3.4, 4.6, 12 and 22 $\mu$m (hereafter
$W1, W2, W3$ and $W4$; \citet{Wright10}). 
The survey was conducted with a 40 cm cryogenically-cooled telescope in sun-synchronous polar orbit. 
Four infrared detectors imaged the same sky field of view during 7.7 s ($W1, W2$) and 8.8 s ($W3, W4$). 
We use here the data of the second-pass processing,
obtained with improved calibration and processing algorithms, superseding those obtained for the preliminary data release.  

From the all aforementioned datasets we present for the first time the analysis of the INTEGRAL, ASAS-SN, KWS and WISE data.
Table~\ref{Tab:phot} gives a summary of the data sources, the bands that were used, and the time span of the observations.

We search for periods using the GLS periodogram \citep{2009A&A...496..577Z}.  
This algorithm uses the principle of the \citet{1976Ap&SS..39..447L} \& \citet{1982ApJ...263..835S} periodograms with some modifications, such as the addition of a displacement in the adjustment of the fit function and the consideration of measurement errors. Compared with the classical periodogram, it gives us more accurate frequencies and a better determination of the amplitudes.  We complement this analysis with 
the $O-C$ method described by \citet{2005ASPC..335....3S}, where predicted times of eclipses of a  test ephemerides are compared with observed times 
in order to get an improved value of the orbital period. 

Furthermore, we disentangled the light curve subtracting from the original light curve a representation of the orbital variability
constructed with a Fourier series including the orbital frequency and its harmonics. This methodology was introduced in  \citet{Mennickent2012}
and used in several past studies of close binaries \citep[e.g.][]{2019MNRAS.487.4169M}. This allowed us to search for new periodicities in the residual data.
The results of this search are given in the next section.

\begin{table}
\caption{Photometric time series analyzed in this paper. KW refers to \citet{2009arXiv0907.1047K}, KWS means  Kamogata Kiso Kyoto Wide-field Survey 
and P02 refers to \citet{2002ARep...46..900P}. N means number of data points and $\sigma$ average data point error.}
\label{Tab:phot}
\centering
\begin{tabular}{l l c c r}
\hline\hline
Source & Band & $\Delta$ HJD-2400000 &N &$\sigma$ (mag)\\
\hline
KW & $U$ & 42655-46055 &955 &NA\\
KW & $B$ & 42655-46055 &955 &NA\\
KW & $V$ & 42655-46055 &955 &NA\\
INTEGRAL & $V$ & (1093-6774)* &3697 &0.013\\
ASAS-SN  & $V$ & 57009-58451&615&0.0058\\
ASAS-SN   & $g$ & 58610-58795&245&0.0025\\
KWS & $V$& 56268-58807 &748&0.027\\
KWS & $I_c$ &  56484-58807 &616&0.027\\
WISE &$W1$&(55216-55411)** &55  &0.028\\
WISE &$W2$&(55216-55411)** &55  &0.021\\
WISE &$W3$&(55216-55411)** &37  &0.023\\
WISE &$W4$&(55216-55411)** & 37 &0.163\\
P02&  $U$ & *** &47 & 0.028\\
P02& $B$ & *** &55 & 0.020\\
P02& $V$ & *** &55 & 0.018\\
P02& $R$ & *** &59 & 0.016\\
\hline       
\end{tabular}
\vspace{0.25cm}\\
* The INTEGRAL satellite data is in Barycentric Julian Date 
(BJD-2451544.5). ** The WISE satellite data is in Mean Julian Date (MJD-2400000).  ***Authors give only phases versus differential magnitudes. 
\end{table}

\begin{figure}[]
\begin{center}
\includegraphics[angle=-90,width=1\linewidth]{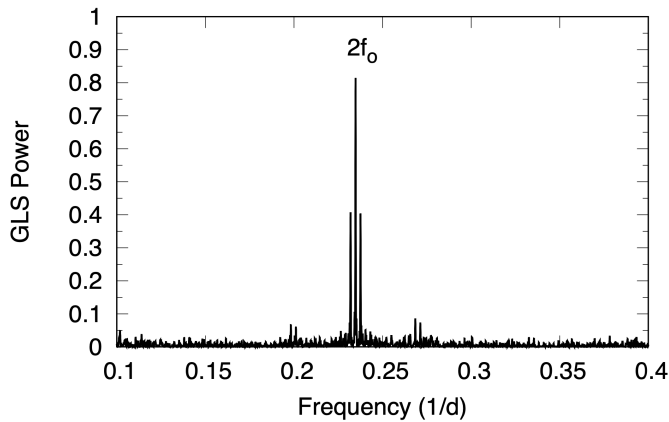}
\end{center}
\caption{Periodogram obtained with the $V$-band photometry. $\rm 2f_{0} = 2/P_{o}$, where $\rm P_{o} = 8\fd519252 \pm 0\fd000012$ d.} 
\label{Operiodogram}
\end{figure}

\begin{figure*}[h!]
\begin{center}
\includegraphics[width=1.0\linewidth]{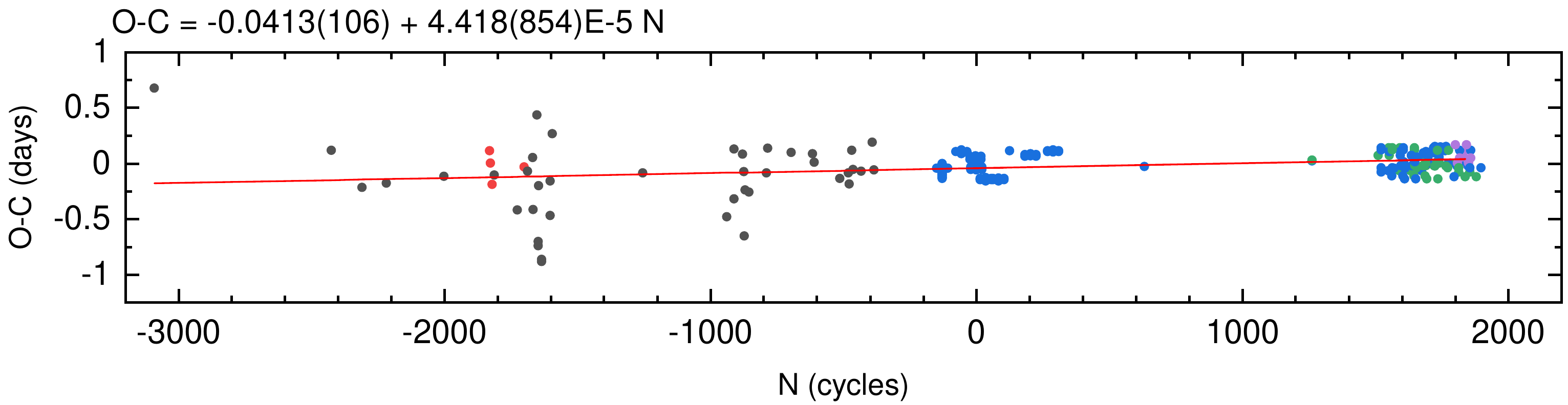}
\end{center}
\caption{Observed (O) minus calculated (C) epochs for primary minima versus cycle number for 117 years of observations, according to the ephemeris given by Eq.\,1.
 Black and red dots indicate photographic and visual data, respectively.  Blue, green and magenta dots show CCD or photoelectric $V$, $I_c$ and $g$-band data, respectively. 
The data points with the same X and Y coordinates are displayed with a small X offset for better visualization. See Section 3.1. } 
\label{ocfit}
\end{figure*}

\begin{figure}[]
\begin{center}
\includegraphics[width=1.0\linewidth]{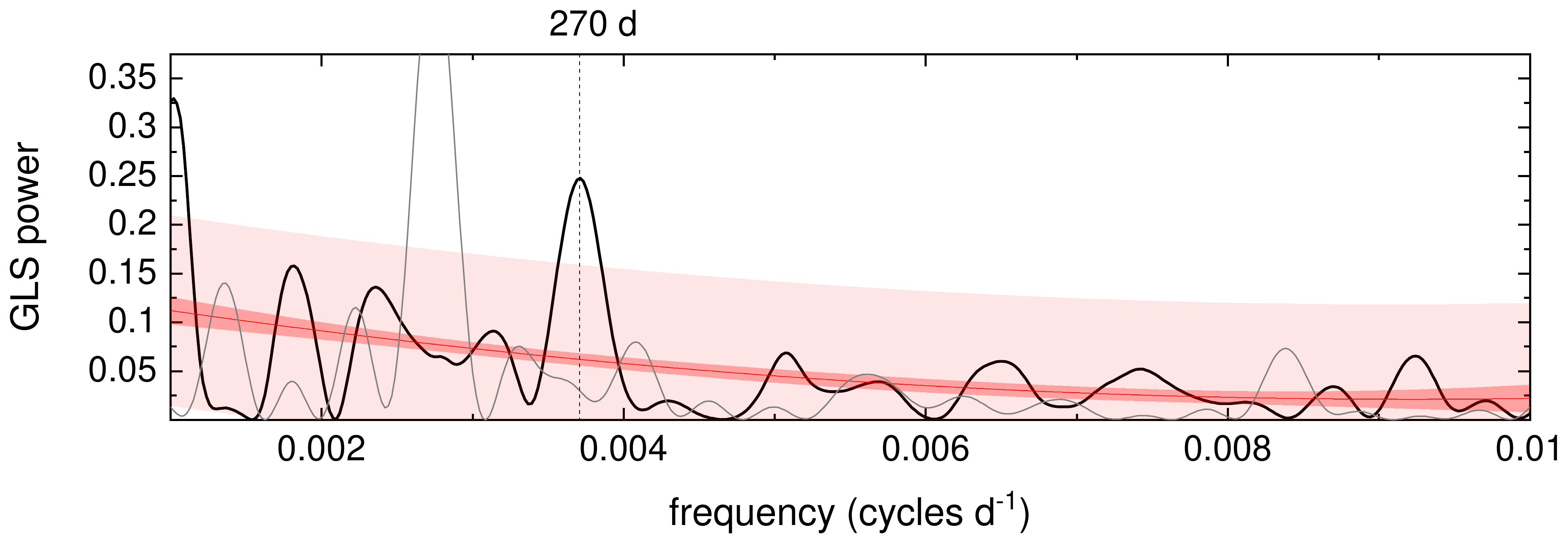}\\
\end{center}
\caption{Generalized Lomb-Scargle periodogram obtained with $I_c$-band data after subtraction of 
the orbital period. The light gray curve shows the spectral window for the data. The best 2$^{th}$ order polynomial fit is also shown.  95\% confidence and prediction bands are shown by dashed and  dashed-light areas, respectively.} 
\label{perlong}
\end{figure}

\begin{figure}
\begin{center}
\includegraphics[width=1.0\linewidth]{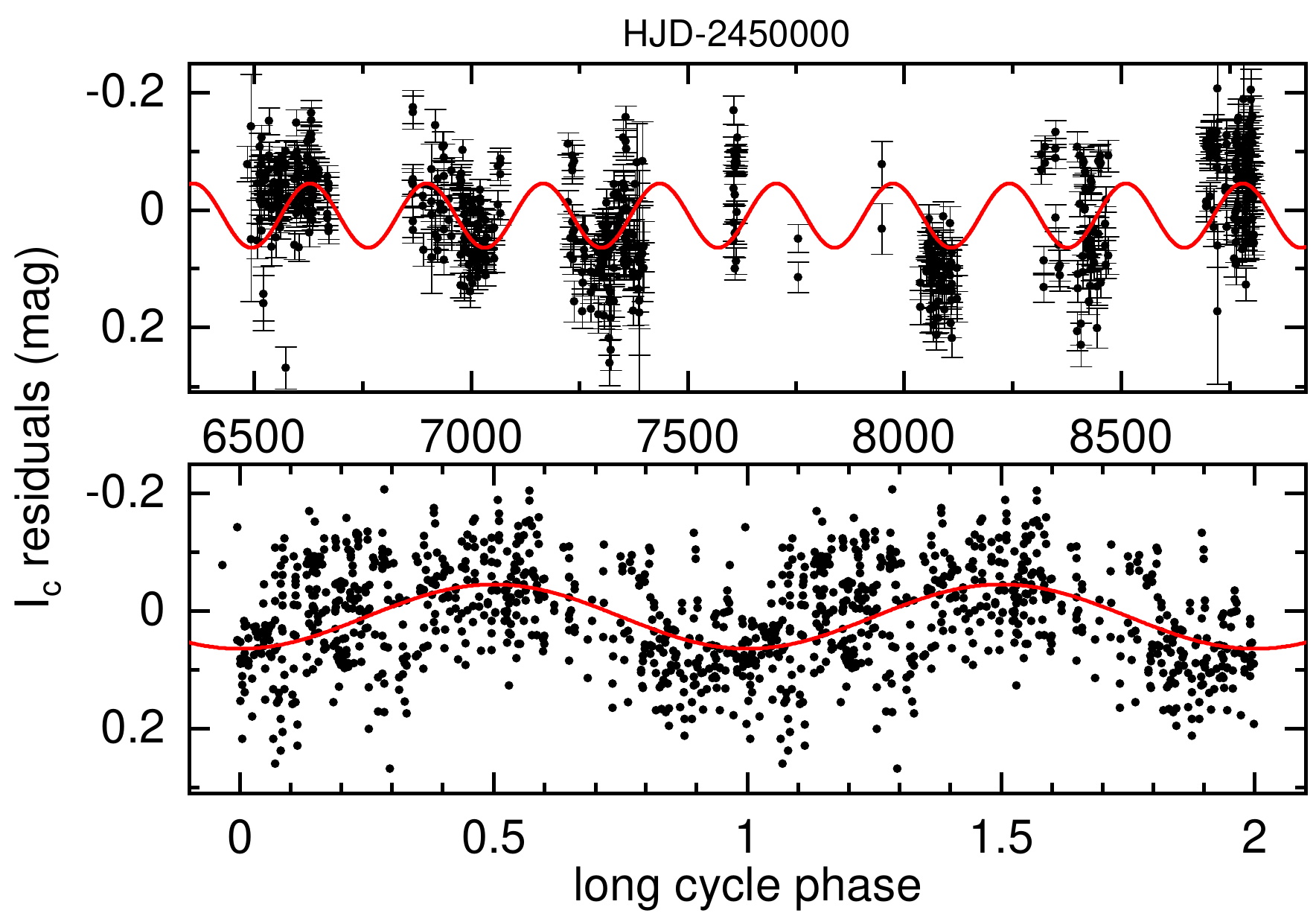}
\end{center}
\caption{ Upper: $I_{c}$-KWS light curve after removing the orbital period. Lower: The same but phased with the long period of 269 days. The best sinus fit is also shown. } 
\label{lcycle}
\end{figure}

\section{Results}
\label{Sec:Results}

\subsection{The orbital period }


We searched for the orbital period 
using the $V$-band photometry, finding an orbital period of $8\fd519252(12)$ (Fig.\ref{Operiodogram}). This value compares well with that provided by 
\citet{2004AcA....54..207K}, viz.\,8\fd519281. The following ephemeris for the primary eclipse was found:

\begin{equation}
  \rm HJD_{min} = 2443132.29826 + E\times 8\fd519252(12)
 \label{eq1}
\end{equation}

In order to improve the accuracy of the orbital period, we selected the photometric data points with phases close to the primary minimum 
($\rm 0.98 \leq \Phi_{o} \leq 1.02$) and performed an analysis of observed (O) minus calculated (C) eclipse times following  \citet{2005ASPC..335....3S}. 
In this analysis the O-C deviations can be represented as a function of the number of cycles with a straight line whose
zero point and slope are the corrections needed for the (linear) ephemeris zero point and period, respectively. We include 48 times of primary minimum 
studied by \citet{2004AcA....54..207K} and listed in the O-C Gateway\footnote{http://var2.astro.cz/ocgate/index.php?lang=en}. Additionally, we include
209 new times of minima measured from the data listed in Table\,1,   excluding satellite data because of the different time system. The dataset of minima covers about 5000 cycles i.e. 
117 years of observations. 
 Since some minima times are published without errors, we use a simple least square fit for our analysis.
Our result displayed in Fig.\ref{ocfit} shows that the fit can be performed with a straight line, indicating a constant period. 
The new ephemeris is given by:

\begin{eqnarray}
 \rm HJD_{min} = 2443132.25696 + E\times 8\fd519296(8)
 \label{eq2}
\end{eqnarray}

\begin{figure*}[]
\begin{center}
\includegraphics[width=1.0\linewidth]{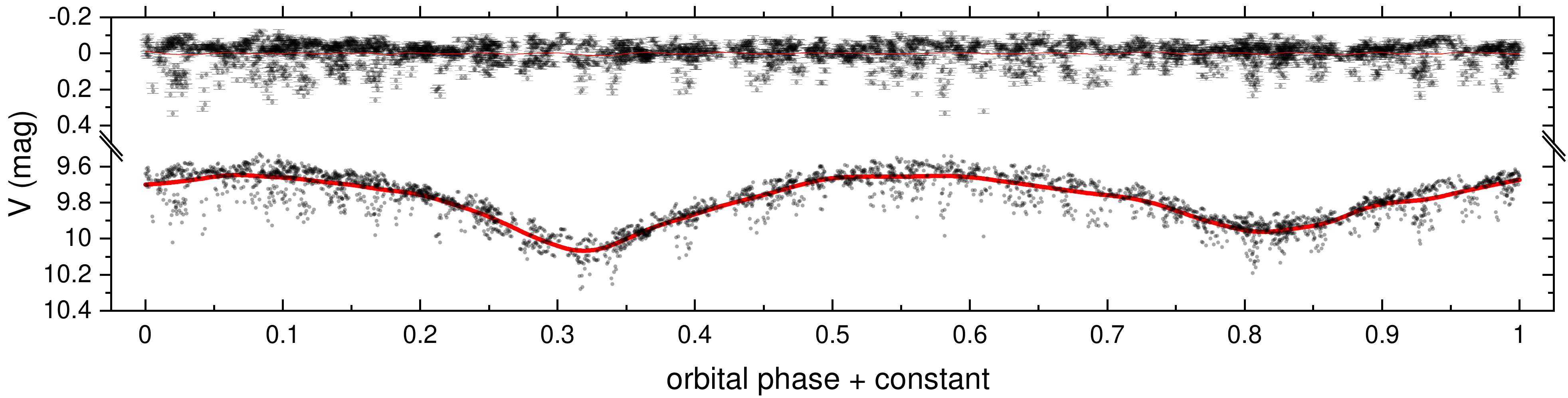}
\end{center}
\caption{Integral $V$-band data phased with the ephemeris given by Eq.\,2, along with a locally weighted scatterplot smoothing curve. 
Residuals are shown in the upper part with data points error bars. The orbital phase has been displaced for better visualization.} 
\label{integral}
\end{figure*}

\begin{figure}[]
\begin{center}
\includegraphics[width=8.5cm]{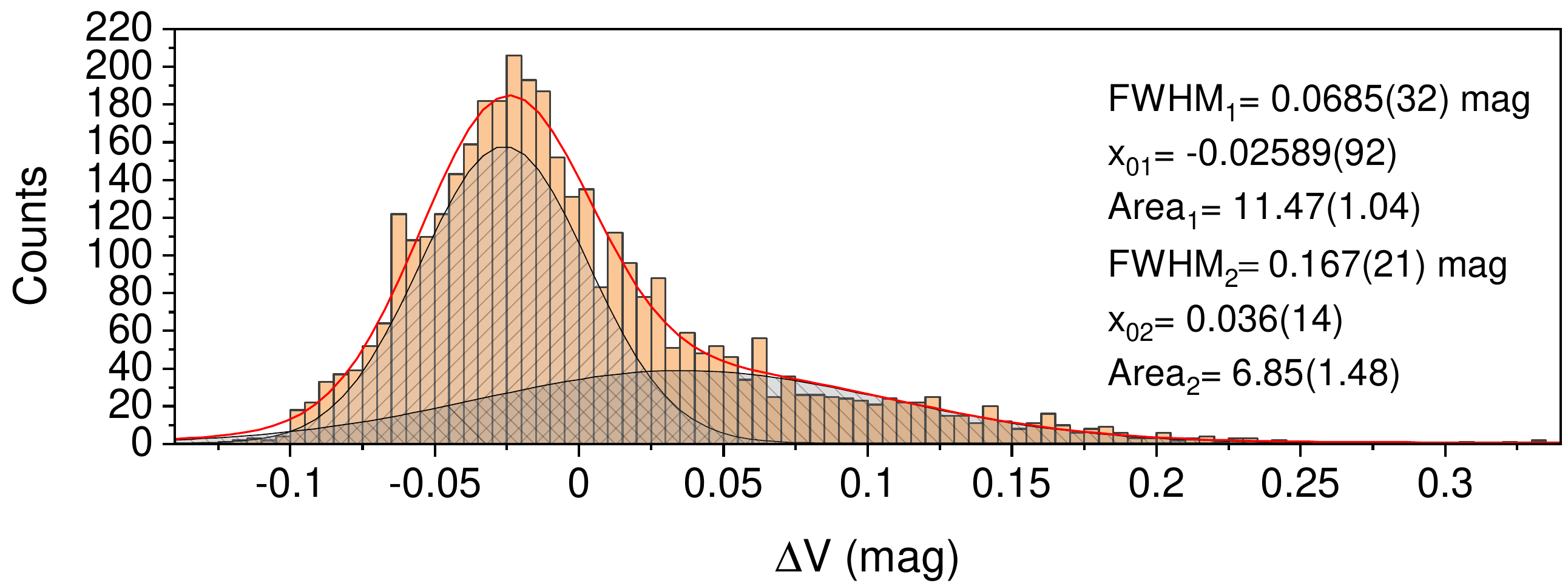}
\end{center}
\caption{Histogram of residuals shown in Fig.\ref{integral} modeled by the sum of two  Gaussians. A  non-linear fit using the 
Levenberg Marquardt iteration algorithm results in two distributions (black lines) and its sum (red line). The  adjusted parameters full width at half maximum, center, area and their errors
are given for the two  Gaussians.
} 
\label{histo}
\end{figure}

\begin{figure}[]
\begin{center}
\includegraphics[width=8.5cm]{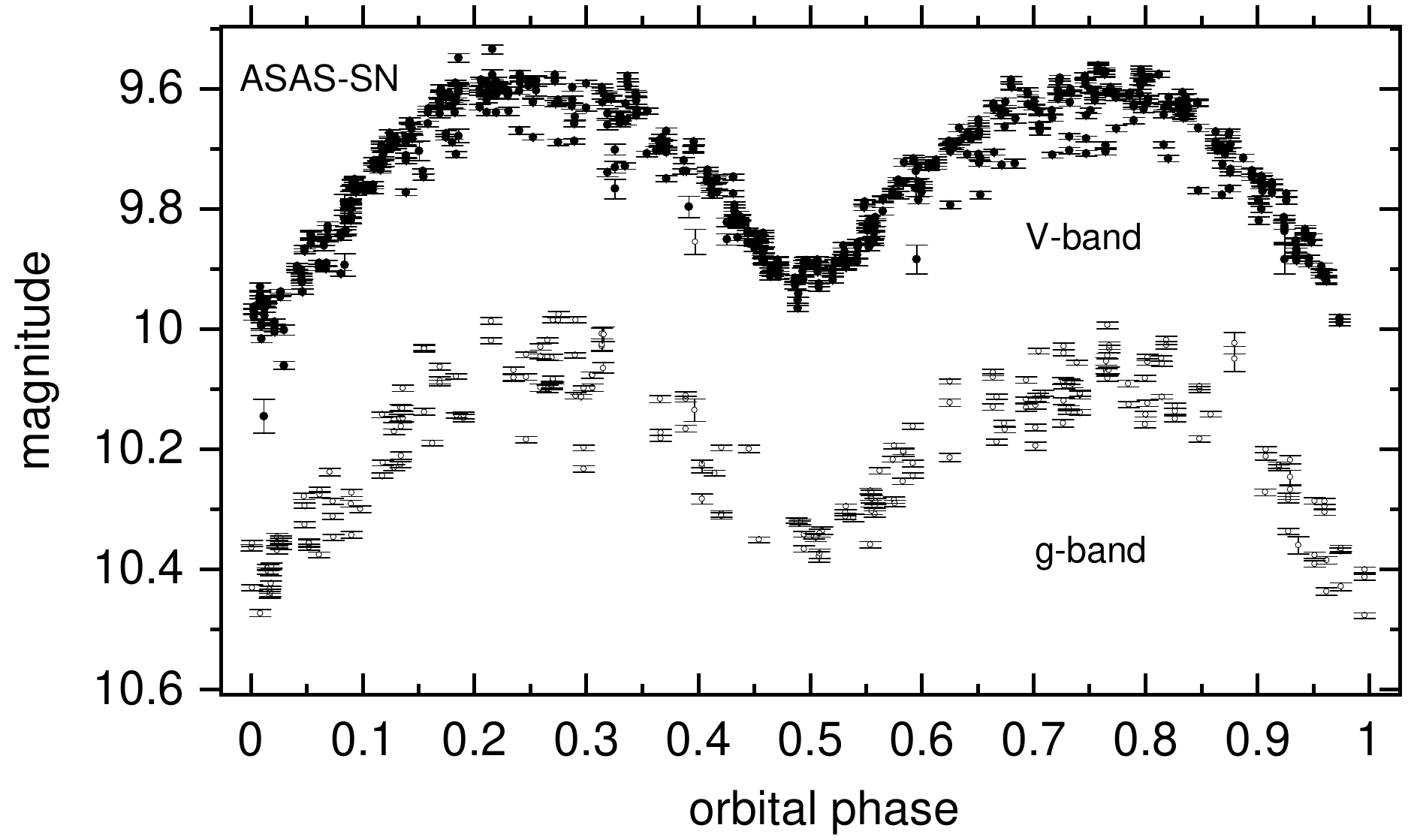}
\end{center}
\caption{ASAS-SN $V$ and $g$-band magnitudes phased with the ephemeris given by Eq.\,2. The drops in brightness seen in INTEGRAL data are also present.  Phase zero corresponds to the primary (deeper) eclipse.}
\label{Vg}
\end{figure}

\subsection{Search for additional photometric periods }

We disentangled the light curve removing the orbital frequency finding a long period in the residuals  of the $I_{c}$-KWS band photometry; 268.7 $\pm$ 1.6 d with  full 
amplitude 0.1096 $\pm$ 0.0066 mag. The following ephemeris  for the minimum of the long-cycle was found: 

\begin{eqnarray}
 \rm HJD_{min} = 2456493.37 + E\times (268\fd7 \pm 1\fd6)
 \label{eq3}
\end{eqnarray}

No periodicity was found in the residuals of the other datasets. 
The periodogram showing the peak at the long cycle frequency is shown in 
Fig.\ref{perlong}. The residual light curve phased with the long cycle is shown in 
Fig.\ref{lcycle}. The peak of the $I_c$-band periodogram is significant  and cannot be explained in terms of the
data sampling. \\

\subsection{Flickering in the light curve}

We observe flickering in the light curve, better represented in the INTEGRAL database (Fig.\ref{integral}). This phenomenon 
consists of drops of brightness in the light curve regarding the mean level for a given orbital phase. It was previously reported as model deviations in several bands up to 0.1 mag
in earlier work with less time-resolved datasets by \citet{2002ARep...46..900P}. Larger variations of up to 0.2 mag during 9 nights between 1972 and 1973 were recorded by \citet{2009arXiv0907.1047K} but interpreted in terms of an instrumental effect. The drops in brightness observed in INTEGRAL data cannot be due to measurements errors:  individual data point errors are much smaller than the observed deviations and the expected instrumental error for each data point for the given system magnitude is less than 0.02 mag  \citep{2003A&A...411L.261M}.

 A Levenberg Marquardt nonlinear curve fit was performed with two Gaussians characterized by free parameters 
position, full width at half maximum and area. The fit converged after 19 iterations with a $\chi^2$ tolerance  value of 10$^{-9}$ and R-square value of 0.977.  
The best fit reveals two different distributions 
separated by 0.062 $\pm$ 0.014 mag and whose heights are in the ratio 4.1 $\pm$ 0.6 (Fig.\ref{histo}). The lower  
distribution is 2.4 $\pm$ 0.3 times broader than the tallest distribution. The areas of the distributions are $A_1$ = 11.5 $\pm$ 1.0 and $A_2$ = 6.8 $\pm$ 1.5 mag $\times$ counts.  The magnitudes in the half of the smaller distribution corresponding to larger residuals (fainter magnitudes) can be identified as fading events. 
The corresponding ratio of areas $\eta$ = $0.5A_2/A_1$ is the relative fraction of magnitudes that can be classified as magnitude drops. We find that 29.6\% of the observations corresponds to fading events in this dataset. In addition, we find  that the drops  occurs randomly,  none periodicity was found in the residuals of INTEGRAL data. 
They also appear in ASAS-SN $V$ and $g$-band data (Fig.\ref{Vg}). 
The events  have not clear dependence on the orbital phase, they appear at all phases, but seems to be more abundant in quadratures. The larger drops amount to 0.3 mag in $V$-band, i.e.
a fall of roughly 30\% of the total flux. 

\begin{figure*}[]
\begin{center}
\includegraphics[width=1.0\linewidth]{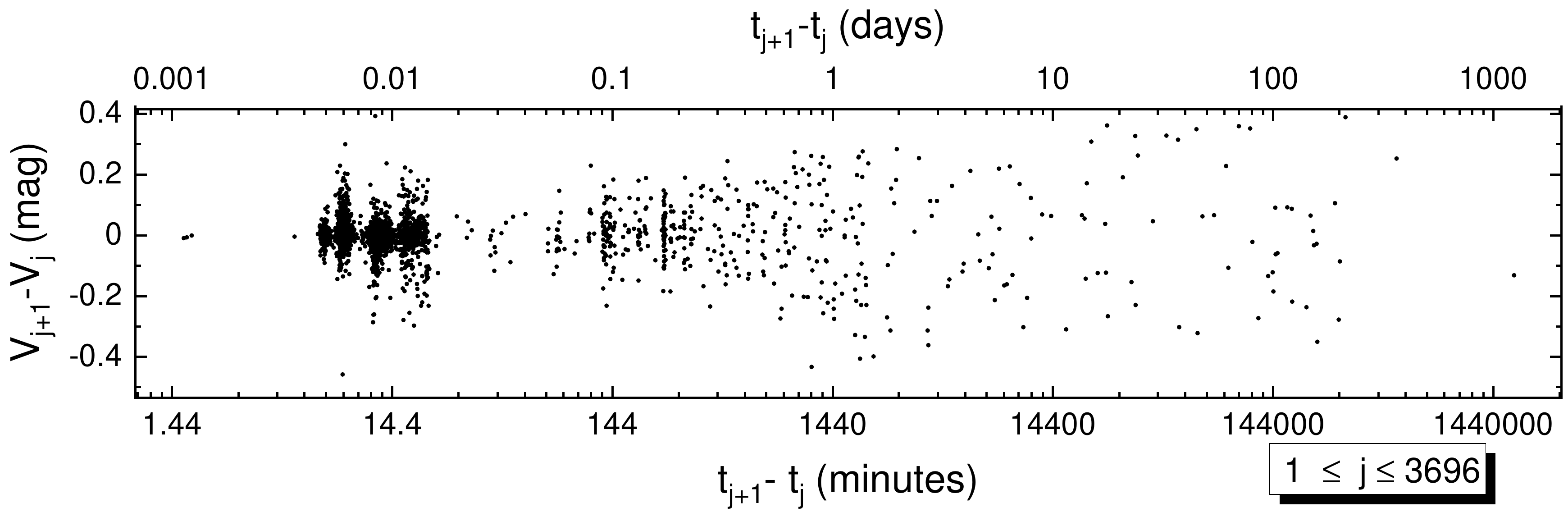}
\end{center}
\caption{
  The temporal distance between adjacent data points of all INTEGRAL $V$-band data is plotted against the corresponding difference of magnitude. We find that flickering occurs in time scales of minutes. Data points errors are $\approx$ 0.01 mag.} 
\label{residuals}
\end{figure*}

 In order to investigate the time scale in which these fading events occur, we performed a correlation analysis measuring the temporal distance between adjacent data points 
and the corresponding difference of magnitude. We find 
that jumps in magnitude of the order of 0.1 mag already occur in time scale of minutes, and conclude that the time resolution of the data is not enough to resolve these events
 (Fig.\ref{residuals}).

\subsection{Infrared colors and $W1$ orbital variability}

\begin{figure*}[]
\begin{center}
\includegraphics[width=0.8\linewidth]{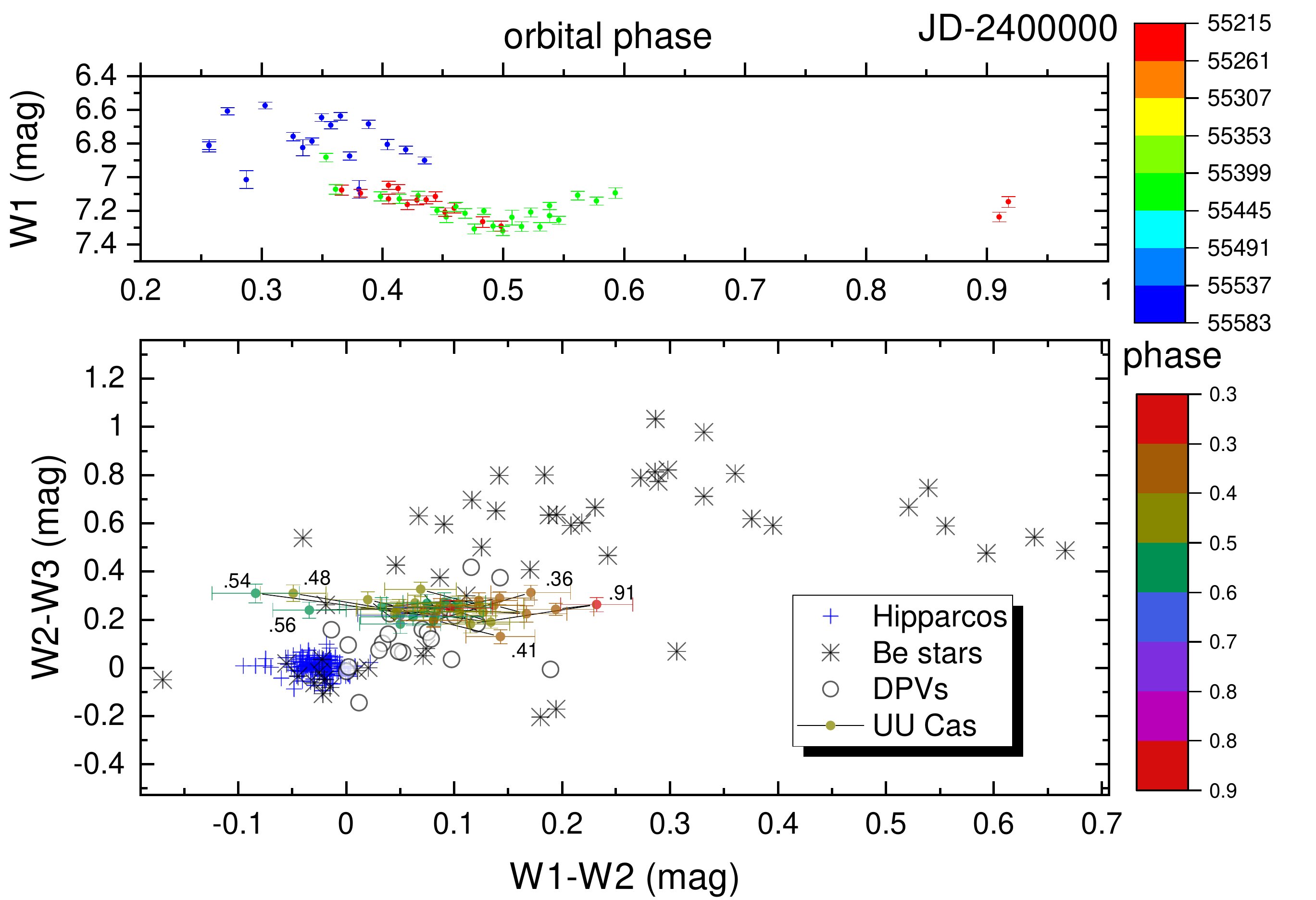}
\end{center}
\caption{Upper: $W$1 magnitude versus orbital phase according to ephemeris given by Eq.\,2. Colors represent time intervals.
Lower: Color-color diagram for non variable HIPPARCOS stars of spectral type B1\,V - K3\,V, Be stars and Double Periodic Variables (DPVs)
according to \citet{2016MNRAS.455.1728M}.  The UU\,Cas data is added with colors  along with numbers (in some cases) showing the orbital phases.   Phase zero corresponds to the primary (deeper) eclipse.} 
\label{wise}
\end{figure*}

We analyzed the $W1$ variability and compared the infrared colors with those of Double Periodic Variables (DPVs) and Be stars. DPVs
are close binaries mostly consisting of a B-type hotter star surrounded by an accretion disk fed by Roche-lobe overflow and mass transfer from a cooler giant companion
\citet{2017SerAJ.194....1M}. Be stars are rapidly rotating B-type stars surrounded by a circumstellar disk. Both types of objects usually show color excess mostly attributed to circumstellar reddening, therefore they are used here for comparison of presence of circumstellar matter. No correction for interstellar reddening was performed. However, at these wavelengths, the effect of interstellar reddening is negligible. In addition, we included as a reference the average colors for 136
main sequence stars in the range of spectral types from B1\,V to K3\,V from the Hipparcos catalogue \citep{1997A&A...323L..49P}.
For details of data selection of DPVs, Be stars and Hipparcos stars see \citet{2016MNRAS.455.1728M}. Our results are shown in Fig.\ref{wise}.

We find that the light $W$1 magnitude follows the orbital period. Our time resolution and cadence does not allow to resolve the whole orbital cycle neither the possible long-cycle.
Part of the variability occurs in longer time scales than the orbital one, this can be seen in the 
data acquired in the end of the time series, that deviate to brighter magnitudes compared with the rest of the data of similar orbital phase. 
In addition, the system appears in the  $W1-W2$ vs. $W2-W3$ diagram in a place populated by systems with circumstellar envelopes, far from the place 
where the HIPPARCOS non-variable stars are located. This finding strongly suggests that circumstellar matter is also present in UU\,Cas.
In addition, during the orbital cycle, the system transits mostly in the $W1-W2$ axis, describing a horizontal path in the color-color diagram.

\begin{figure}[]
\begin{center}
\includegraphics[width=0.9\linewidth]{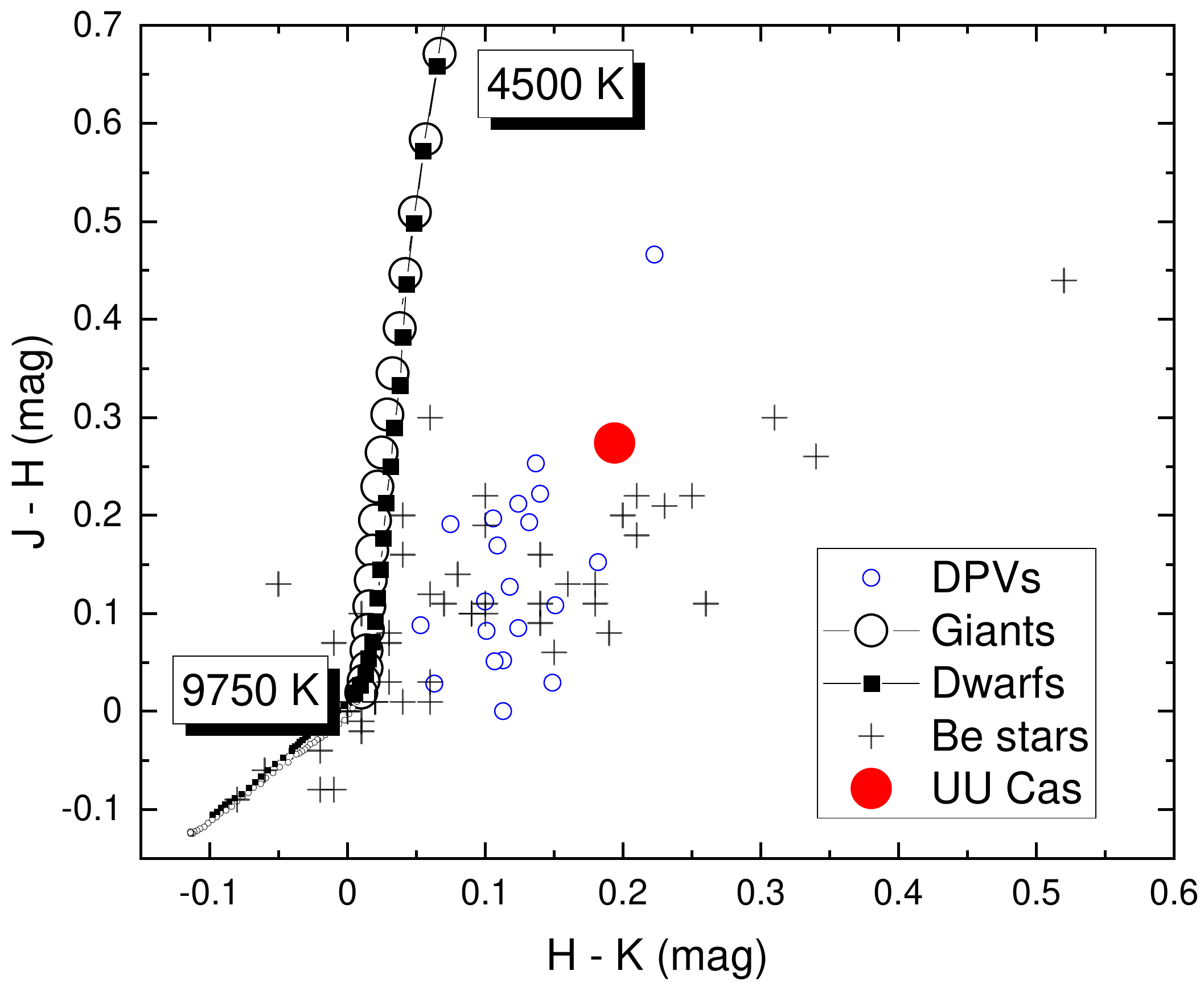}
\end{center}
\caption{Colors for Galactic DPVs \citep{2016MNRAS.455.1728M}, Be stars \citep{2001A&A...369...99H} and synthetic stellar atmosphere models with \logg\ = 4.0 (squares with lines) 
and \logg\ = 3.0 (open circles with lines) from \citet{1998A&A...333..231B}. The effective temperature of 2 selected synthetic models have been labeled along with the position of UU\,Cas.} 
\label{jhk}
\end{figure}

The same tendency is found when comparing the published 2MASS  $J-H$ (0.194) versus $H-K$ (0.274) colors of UU\,Cas \citep{2006AJ....131.1163S} with those of synthetic models for dwarfs and giants, along with 
observed colors of Be stars and DPVs (Fig.\ref{jhk}). The UU\,Cas system is located in a similar place regarding the Double Periodic Variables, suggesting the existence of circumstellar matter and eventually an accretion disk.

\section{Light curve model}

In this section we model the orbital light curves with a theoretical code that solves the inverse problem considering a semi-detached 
system consisting of a  less massive and cooler "donor"  star and a more massive and hotter "gainer" star surrounded by an accretion disk, that is both optically and geometrically thick
\citep{1992Ap&SS.196..267D, 1992Ap&SS.197.17D, 1996Ap&SS.240..317D}. The model includes hot and bright spots in the disk, following evidence found in previous observations of algols \citep{2004AN....325..229R}. These active regions influence the shape of the light curve during the ingress and egress of the eclipses. The temperature of the disk matches the 
gainer's one in its inner edge, and decreases with a radial profile described by an exponent $a_T$. The model has been described in several papers, so we remit the interested reader to \citet{2013MNRAS.432..799M, bib:paper6, 2019MNRAS.487.4169M}.

We fix the donor temperature  and mass ratio to $\rm T_{c}$ = 22\,700\,K and $q$= 0.52, following results from the spectroscopic studies by \citet{2019ApJ...883..186K} and \citet{2017AstBu..72..321G}
and also our own work based on disentangling of spectra (work in preparation).
We also assume synchronous rotation for the  donor, as expected for a close binary that rapidly synchronize stellar spins with the orbit due to tidal forces.  
On the other hand, the gainer might have been spun-up to a high rotation due to nearly tangentially infalling material \citep{1981A&A...102...17P}, 
hence we have assumed critical rotation for it. For $I_c$ data, we modeled the residuals after subtracting the long cycle.

\subsection{Results of the light curve model}

\begin{figure*}[h!]
\begin{center}
\includegraphics[width=0.8\linewidth]{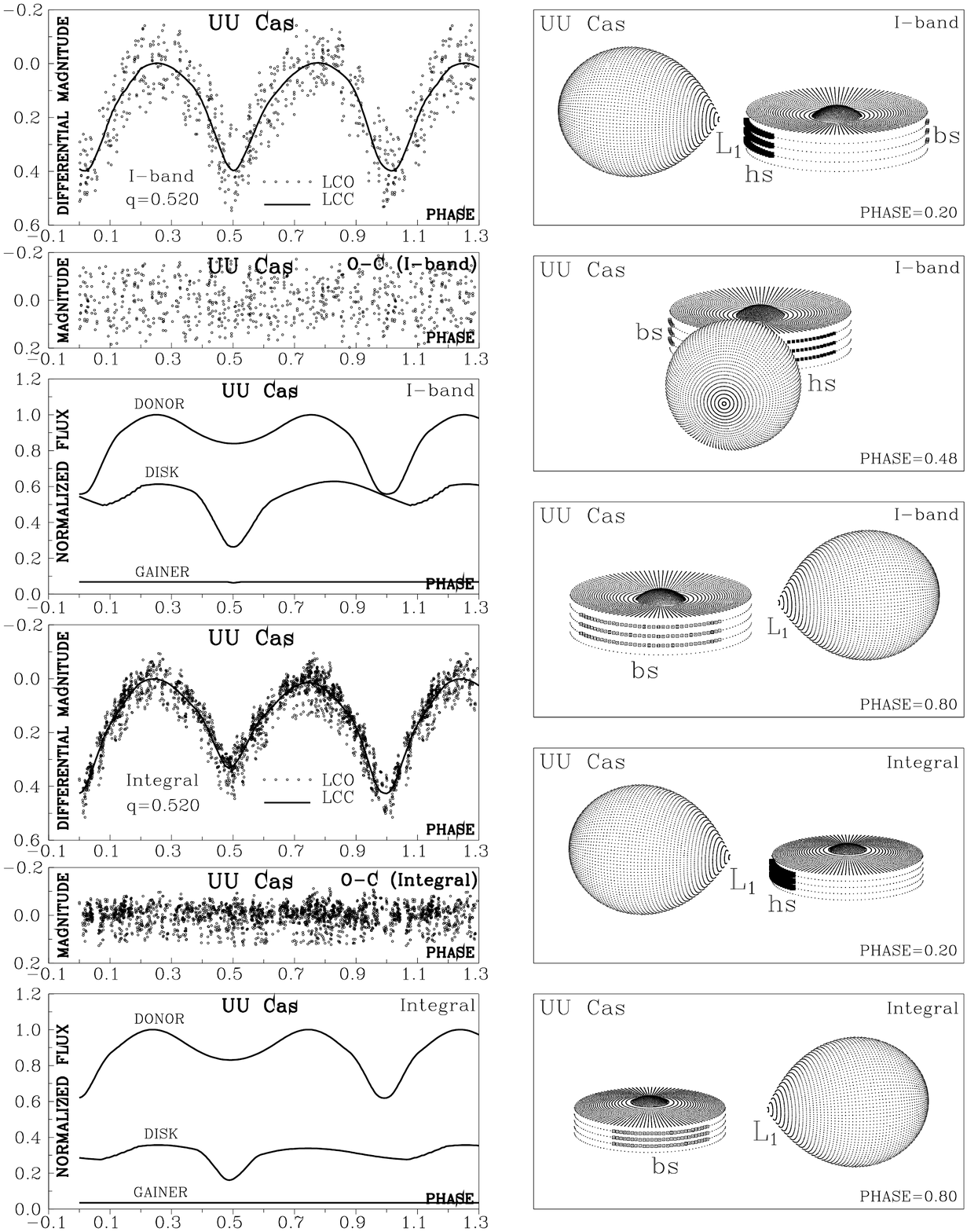}
\caption{Observed (LCO) and synthetic (LCC) light-curves of
{$\rm UU \ Cas$} obtained by analyzing {\rm KWS-$I_c$} data and INTEGRAL  $V$-band  photometric observations; final
O-C residuals between the observed and optimum synthetic light curves;
fluxes of donor, gainer and of the accretion disk, normalized to the donor flux
at phase 0.25; the views of the optimal model at orbital phases 0.20, 0.48 and 0.80
({\rm $I_c$-KWS})  and  0.20 and 0.80 (INTEGRAL), obtained with parameters estimated by the light curve analysis.}
\label{models}
\end{center}
\end{figure*}

The parameters of the best fits are shown in Table\, \ref{TabUUCas}. 
 KWV-$UBV$ and P02-$UBVR$ multi-color data was fit independently and parameters were averaged.
Examples of fits to the KWS-$I_c$ and Integral $V$-band light curves are shown in Fig.\ref{models}. We obtain a good match between observed and calculated magnitudes in all studied cases,
along with consistency  for the stellar and orbital parameters. The disk parameters show some variations, especially the bright spot position, the spots temperature and the 
disk vertical thickness and radius. 
Our model indicates a system seen at angle 74\dg\ and a stellar separation of 52 ${\rm R_{\odot}}$. The stellar temperatures are 22\,700\,$K$ and 30\,200\,$K$. 
The stellar masses are 9 and 17.4 ${\rm M_{\odot}}$ and surface gravities \logg\ =  2.94 and 3.98. The stellar radii are 16.9 and  7.0 ${\rm R_{\odot}}$. The stellar 
temperatures indicate spectral types of B\,2 and B\,0  for the donor and gainer stars \citep{1988BAICz..39..329H}.

The disk radius is relatively stable around a value of 21 ${\rm R_{\odot}}$, reaching its smaller value at 18 ${\rm R_{\odot}}$ (INTEGRAL-$V$ data) and its maximum value at 22 ${\rm R_{\odot}}$ (P02 data). It has a temperature of 16\,000\,$K$ at its outer edge. 
The inner disk thickness has an average value of 9 ${\rm R_{\odot}}$ and the outer vertical thickness has an average  value of 6.5 ${\rm R_{\odot}}$, being maximum at 9.4 ${\rm R_{\odot}}$
(KWS-$I_c$ data).
The hot spot is 36\% (P02 data) to 70\% (KWS-$I_c$ data) hotter than the disk and is located consistently 42\dg\ apart from the line joining the stellar centers in direction opposite to the orbital motion,
i.e. it  is located roughly at the expected region where the gas stream impacts the disk.  The bright spot is 24\% (P02 data) to 50\% (KWS-$I_c$ data) hotter than the disk and changes its location considerably, being located between 82\dg (KW data) and 141\dg (KWS-$g$ data) from the line joining the stellar centers in direction of the orbital motion. 
 At quadratures, the flux ratio between donor and disk  is $\approx$ 2 in the $I_c$-band,  while in the $V$-band this ratio is $\approx$ 3.
The gainer is  almost completely hidden by the disk at all orbital phases contributing less than 10\% to the total flux in both bands.

\begin{table*}[h!]
\small
\caption{Results of the analysis of  {$\rm UU \ Cas$} orbital light curves.
Dataset labels refers to Table\,1. 
The models are obtained by solving the inverse problem for the Roche model with
an accretion disk around the more-massive (hotter) gainer in critical
non-synchronous rotation regime \citep{1992Ap&SS.196..267D, 1992Ap&SS.197.17D, 1996Ap&SS.240..317D}. Mean values are also given.}
 \label{TabUUCas}

        \[
        \begin{array}{lllllllll}

               \hline

               \noalign{\smallskip}
{\rm Quantity}&UBV& UBVR& g& V& I_c& V &{\rm Average \pm std}\\
{\rm  }&{\rm  KW}& {\rm  P02}& {\rm  ASAS-SN}& {\rm  KWS}& {\rm  KWS}&  {\rm INTEGRAL} &\\
            \noalign{\smallskip}

            \hline

            \noalign{\smallskip}

   q                               & 0.52           & 0.52          & 0.52         & 0.52        & 0.52        & 0.52      &0.52  \\
   i {\rm [^{\circ}]}              & 74.2 \pm 0.3   & 74.7 \pm 0.3  & 74.3 \pm 0.6 &74.3 \pm 0.4 &74.9 \pm 0.4 &74.6 \pm 0.5 &74.5 \pm 0.3\\
{\rm F_d}                          & 0.91 \pm 0.04  & 0.97 \pm 0.03 & 0.95 \pm 0.03& 0.88\pm 0.02&0.96 \pm 0.03&0.80 \pm 0.03&0.91 \pm 0.06\\
{\rm T_d} [{\rm K}]                & 16180\pm 390   & 14970\pm 300  &16030 \pm 300 &16320\pm 300 &16100\pm 300 &15720 \pm 300&15887 \pm 491\\
{\rm d_e} [a_{\rm orb}]            & 0.116\pm 0.006 & 0.138\pm 0.03 &0.082 \pm 0.01&0.109\pm 0.02&0.179\pm 0.01&0.121 \pm 0.02&0.12 \pm 0.03\\
{\rm d_c} [a_{\rm orb}]            & 0.171\pm 0.016 & 0.178\pm 0.01 &0.190 \pm 0.01&0.183\pm 0.01&0.107\pm0.01&0.189\pm 0.01&0.17 \pm 0.03\\
{\rm a_T}                          & 8.5  \pm 0.6   & 8.0  \pm 1.4  & 8.3  \pm 0.9 & 8.4 \pm 0.6 &8.5 \pm 0.6 &8.2  \pm 0.5 &8.3 \pm 0.2\\
{\rm f_g}                          & 11.8 \pm 0.2   &11.95 \pm 0.1  &11.91 \pm 0.2 &11.97\pm 0.2&12.0 \pm 0.3 &12.0 \pm 0.2 &11.94 \pm 0.08\\
{\rm F_h}                          & 1.00           & 1.00          & 1.00         & 1.00        & 1.00        & 1.00       &1.00 \\
{\rm T_h} [{\rm K}]                       & 30260\pm 400   & 30150\pm 400  & 30260\pm 400 &30230 \pm 400&30200 \pm 400&30220\pm 400 &30220 \pm 42 \\
{\rm T_c} [{\rm K}]                       &  22700          & 22700         & 22700        & 22700       & 22700       & 22700      &22700 \\
{\rm A_{hs}=T_{hs}/T_d}             & 1.46 \pm 0.13  & 1.36 \pm 0.12 & 1.50 \pm 0.08&1.57 \pm 0.07& 1.70\pm 0.06&1.54 \pm 0.07&1.52 \pm 0.11\\
{\rm \theta_{hs}}{\rm [^{\circ}]}    & 22.8 \pm 1.2   & 24.4 \pm 0.9  & 22.9 \pm 1.2 &22.6 \pm 1.0 &22.8 \pm 0.9 &23.2 \pm 1.1& 23.1	\pm 0.7\\
{\rm \lambda_{hs}}{\rm [^{\circ}]} & 313.0\pm 5.1   & 316.5\pm 12.8 &326.0 \pm 7.1 &313.5\pm 5.6 &320.5\pm 6.2 &319.1\pm 7.1&318.1 \pm	4.9 \\
{\rm \theta_{rad}}{\rm [^{\circ}]}   & -31.0\pm 6.1   & -30.2\pm 5.1  &-27.4 \pm 5.6 &-29.3\pm 7.0 &-31.8\pm 8.2 &-30.9\pm 9.2&-30.1	\pm 1.6 \\
{\rm A_{bs}=T_{bs}/T_d}             & 1.31 \pm 0.05  & 1.24 \pm 0.08 & 1.40 \pm 0.08&1.33 \pm 0.07&1.50 \pm 0.08&1.34 \pm 0.07&1.35 \pm 0.09\\
{\rm \theta_{bs}} {\rm [^{\circ}]}   & 52.4 \pm 3.1   & 51.8 \pm 2.2  & 52.0 \pm 2.7 &51.6 \pm 3.6 &52.5 \pm 4.0 &40.5 \pm 3.2 &50.1 \pm 4.7\\
{\rm \lambda_{bs}}{\rm [^{\circ}]} & 82.2 \pm 13.2  &124.9 \pm 12.9 & 141.2\pm 9.7 &107.9\pm 8.9 &118.4\pm 9.4 &98.2 \pm 7.5&112.1 \pm 20.8 \\
{\Omega_{\rm h}}                        & 9.467\pm 0.03  & 9.554\pm 0.02 & 9.534\pm 0.06&9.562\pm 0.05&9.592\pm 0.06&9.586\pm 0.05&9.55	\pm 0.04\\
{\Omega_{\rm c}}                        & 2.914\pm 0.02  & 2.914\pm 0.02 & 2.914\pm 0.02&2.914\pm 0.02&2.914\pm 0.02&2.914\pm 0.02&2.914 \pm  0.02\\
\cal M_{\rm_h} {[\rm M_{\odot}]}  & 17.4 \pm 0.2   & 17.4 \pm 0.2  & 17.4 \pm 0.3 &17.4 \pm 0.3 &17.4 \pm 0.3 &17.4 \pm 0.3&17.4 \pm  0.3 \\
\cal M_{\rm_c} {[\rm M_{\odot}]}  & 9.0  \pm 0.2   & 9.0  \pm 0.2  & 9.0  \pm 0.2 & 9.0 \pm 0.2 &9.0 \pm 0.2 &9.0 \pm 0.2 &9.0 \pm  0.2\\
\cal R_{\rm_h} {\rm [R_{\odot}]}   &  7.1 \pm 0.1   & 7.0  \pm 0.1 & 7.0  \pm 0.1 & 7.0 \pm 0.1& 7.0 \pm 0.1&7.0 \pm 0.1& 7.0 \pm  0.1\\
\cal R_{\rm_c} {\rm [R_{\odot}]}   & 16.9 \pm 0.1   & 16.9 \pm 0.1  & 16.9 \pm 0.1&16.9 \pm 0.1&16.9\pm 0.1&16.9\pm 0.1& 16.9 \pm  0.1\\
{\rm log} \ g_{\rm_h}                    & 3.98 \pm 0.02  & 3.98 \pm 0.02 & 3.98 \pm 0.02&3.98 \pm 0.02&3.99 \pm 0.02&3.99 \pm 0.02&3.983 \pm 0.005\\
{\rm log} \ g_{\rm_c}                    & 2.94 \pm 0.02  & 2.94 \pm 0.02 & 2.94 \pm 0.02&2.94 \pm 0.02&2.94 \pm 0.02&2.94 \pm 0.02&2.94 \pm  0.02\\
M^{\rm h}_{\rm bol}                     &-6.7  \pm 0.1   &-6.6 \pm 0.1   &-6.6  \pm 0.1 &-6.6 \pm 0.1 &-6.6\pm 0.1 &-6.6\pm 0.1&-6.62 \pm 0.04\\
M^{\rm c}_{\rm bol}                     &-7.3  \pm 0.1   &-7.3 \pm 0.1   &-7.3  \pm 0.1 &-7.3 \pm 0.1 &-7.3 \pm 0.1 &-7.3 \pm 0.1&-7.3 \pm  0.1\\
a_{\rm orb}  {\rm [R_{\odot}]}     & 52.2 \pm 0.3   & 52.2 \pm 0.2  &52.2  \pm 0.3 &52.2 \pm 0.3 &52.2 \pm 0.3 &52.2 \pm 0.3&52.2 \pm  0.3 \\
\cal{R}_{\rm d} {\rm [R_{\odot}]} & 20.7 \pm 1.1   & 22.0 \pm 0.6  & 21.6 \pm 0.3 &20.1 \pm 0.3 &21.9 \pm 0.4 &18.3 \pm 0.4 &20.8 \pm 1.4\\
\rm{d_e}  {\rm [R_{\odot}]}        & 6.0  \pm 0.4   & 7.2  \pm 1.2  & 4.3  \pm 0.7 & 5.7 \pm 0.5 &9.4  \pm 0.4 &6.3  \pm 0.4 &6.5 \pm 1.7\\
\rm{d_c}  {\rm [R_{\odot}]}        & 8.9  \pm 0.9   & 9.3  \pm 0.5  & 9.9  \pm 0.7 & 9.6 \pm 0.7 &5.6  \pm 0.6 &9.8  \pm 0.5&8.8 \pm 1.6 \\

\noalign{\smallskip}
            \hline
         \end{array}
      \]

FIXED PARAMETERS: $q={\cal M}_{\rm c}/{\cal M}_{\rm
h}=0.52$ - mass ratio of the components, ${\rm T_c=22\,700 K}$  - temperature
of the less-massive (cooler) donor, ${\rm F_c}=1.0$ -
filling factor for the critical Roche lobe of the donor,
$f{\rm _{c}}=1.00$ - non-synchronous rotation coefficients
of the donor, ${\rm F_h}=R_h/R_{zc}$ - filling factor for the
critical non-synchronous lobe of the hotter, more-massive gainer (ratio of the
stellar polar radius to the critical Roche lobe
radius along z-axis for a star in critical non-synchronous rotation regime) ,
${\rm \beta_{h,c}=0.25}$
- gravity-darkening coefficients of the components, ${\rm
A_{h,c}=1.0}$  - albedo coefficients of the components.

\smallskip \noindent Note: Origin of the photometric observations,
photometric band, $q$ - mass ratio of the components, $i$ - orbit
inclination (in arc degrees), ${\rm F_d=R_d/R_{yc}}$ - disk
dimension factor (the ratio of the disk radius to the critical Roche
lobe radius along y-axis), ${\rm T_d}$ - disk-edge temperature,
$\rm{d_e}$, $\rm{d_c}$,  - disk thicknesses (at the edge and at
the center of the disk, respectively) in the units of the distance
between the components, $a_{\rm T}$ - disk temperature
distribution coefficient, $f{\rm _g}$ - non-synchronous rotation coefficient
of the more massive gainer (in the synchronous rotation regime),
 ${\rm T_h}$  - temperature of the more-massive (hotter) gainer,
${\rm A_{hs,bs}=T_{hs,bs}/T_d}$ - hot and bright spots' temperature
coefficients, ${\rm \theta_{hs,bs}}$ and ${\rm \lambda_{hs,bs}}$ -
spots' angular dimensions and longitudes (in arc degrees, the longitude
is measured in direction of the orbital motion), ${\rm
\theta_{rad}}$ - angle between the line perpendicular to the local
disk edge surface and the direction of the hot-spot maximum
radiation, ${\Omega_{\rm h,c}}$ - dimensionless surface potentials
of the hotter gainer and cooler donor, $\cal M_{\rm_{h,c}} {[\cal
M_{\odot}]}$, $\cal R_{\rm_{h,c}} {\rm [R_{\odot}]}$ - stellar
masses and mean radii of stars in solar units, ${\rm log} \
g_{\rm_{h,c}}$ - logarithm (base 10) of the system components
effective gravity, $M^{\rm {h,c}}_{\rm bol}$ - absolute stellar
bolometric magnitudes, $a_{\rm orb}$ ${\rm [R_{\odot}]}$,
$\cal{R}_{\rm d} {\rm [R_{\odot}]}$, $\rm{d_e} {\rm [R_{\odot}]}$,
$\rm{d_c} {\rm [R_{\odot}]}$ - orbital semi-major axis, disk
radius and disk thicknesses at its edge and center, respectively,
given in solar units.

\end{table*}

\section{Discussion}

Our results confirm the picture summarized by \citet{2017AstBu..72..321G} and \citet{2019ApJ...883..186K}
of a semidetached binary where the less massive star feeds an accretion disk located around the more massive and partly hidden star.
Our light curve models based on 11 different datasets obtained at different epochs and with different photometric bands, indicate consistent results for the overall 
system configuration, especially the orbital and stellar parameters, consolidating the overall view of a massive interacting binary of early B-type components.   

Our light curve models indicate a system of B0\,IV + B2\,III stars with a relatively large total 
mass of about 26 ${\rm M_{\odot}}$. Our stellar masses of 17.4 and 9 ${\rm M_{\odot}}$ are  relatively close to those reported by \citet{2017AstBu..72..321G},
namely 17.7 and 9.5 ${\rm M_{\odot}}$. The mass, radius and temperature of the more massive component fit well the relationships for B-type dwarfs reported by \citet{1988BAICz..39..329H}. However, the less massive component is much larger than expected for its temperature and indicates an evolved star, consistent with its lower surface gravity.   
In addition, we find an accretion disk  around the more massive star and derive for the first time their properties.  
The average disk radius of  21  ${\rm R_{\odot}}$ means ${\rm R_{d}/a}$ $\approx$ 0.40, i.e. the disk outer border is just below the tidal radius for the given mass ratio \citep{1977ApJ...216..822P, 1995CAS....28.....W}. The disk radius shows small changes during the observing epochs, while the disk is characterized by two hot and active photometrically variable regions. 
The changes in these regions (position, extension, temperature) might indicate changes in mass transfer rate modulating the mass flows and disk properties. 
The infrared excess obtained in our photometric study confirms the existence of circumstellar matter while the enigmatic obscuration events mostly visible in INTEGRAL $V$-data 
might be related to inhomogeneities present in the flow of mass loss.  The existence of these fading events suggests a mass flow structure  more complex than provided by our simple model of disk plus two spots. In the future, new data of better accuracy could add new constraints to improve the understanding of the system.

The long period observed in the $I_c$-band and the light curve morphology suggest that UU\,Cas might be a Double Periodic Variable \citep[DPV,][]{2003A&A...399L..47M, 2017SerAJ.194....1M}, a subtype of algols showing $\beta$-Lyrae type light curves that have  an accretion disk surrounding a B-type gainer.  DPVs show long cycles of larger amplitude at red passbands \citep{2010ASPC..435..357M}. This might indicate that the long cycle arises from a light source emitting mostly
at redder wavelengths than the other sources in the system. This could explain why the long cycle is detected (with small amplitude) only in the $I_c$-band in UU\,Cas. All DPVs (except $\beta$ Lyrae) show a constant orbital period
in spite of the evidence of active mass exchange and occasional mass outflows. All these credentials are similar to those of UU\,Cas. In addition, the ratio between the long cycle length and the orbital period is 31.7 for UU\,Cas, 
 relatively close to the average $\sim$ 33 for DPVs.  Since the long period is observed only in one dataset, we suggest confirmation with additional data before claiming that UU\,Cas is a DPV. If confirmed, UU\,Cas should be the  most massive DPV;  typically they have total masses just around 10 ${\rm M_{\odot}}$.

The reported evidence for an H$\alpha$ P-Cygni like profile led \citet{2017AstBu..72..321G} to conclude that a stellar wind emerges from the system. This might indicate that the system
is loosing mass. However, the constancy of the orbital period through one century suggests that this mass loss should be  mild, i.e. the angular momentum of the system remains almost unaltered,
as well as the orbital period. This is consistent with a binary in a relatively stable and mild mass transfer stage.

The error in the orbital period $\epsilon$ = 0\fd000008, given by Eq.\,2, might be interpreted as a possible drift of the orbital period between 
P$_o$ - $\epsilon$ and P$_o$ + $\epsilon$ in $\Delta$t =117 years. The above implies a possible change in the orbital period roughly of less than 2$\epsilon_P$/$\Delta$t = 1.37 $\times$ 10$^{-7}$ d/yr. 
For conservative mass transfer this imposes an upper limit for the mass transfer rate. The expected period change in the conservative case is \citep{1963ApJ...138..471H}:

\begin{eqnarray}
\rm \frac{\dot{P}_o}{P_o} = 3 \dot{\cal M_{\rm_c}} \left( \frac{1}{\cal M_{\rm_{c}}}-\frac{1}{\cal M_{\rm_{h}}} \right)
\end{eqnarray}

\noindent
Using the aforementioned numbers and the derived stellar masses we determine $\dot{\cal M_{\rm_c}}$ $<$ 1.0 $\times$ 10$^{-7}$  $\rm M_{\odot}$/yr.

\section{Conclusions}

Based on the analysis of multi-wavelength light curves and published spectroscopic results we have arrived to the following conclusions:

\begin{itemize}

\item We find an improved orbital period  of $\rm P_{o} =  8\fd519296(8)$ that seems to be stable during the last century. This suggests that the rate of system  mass loss or mass exchange between the stars should be small. 
\item We find a long cycle of length $T$ $\sim$ 270 d in the $I_c$-band data.
\item We find a system seen at angle 74\dg\  with a stellar separation of 52 ${\rm R_{\odot}}$. 
The stellar temperatures are 22\,700\,$K$ and 30\,200\,$K$, masses 9 and 17.4 ${\rm M_{\odot}}$, radii 16.9 and 7.0 ${\rm R_{\odot}}$ 
and surface gravities \logg\ = 2.94 and 3.98.
\item The system follows an horizontal path in the $W1-W2$ vs. $W2-W3$ diagram and is located in the region of systems with circumstellar matter like DPVs and Be stars. 
The location in the $J-H$ vs. $H-K$ diagram also suggests infrared excess due to circumstellar matter.  
\item We find
an accretion disk around the hotter star, with a radius of 21 ${\rm R_{\odot}}$. Two active regions are found, one located roughly in the expected position where
the stream impacts the disk and the other one in the opposite side of the disk. Changes are observed in the disk and spot parameters at different datasets; this might indicate variable mass transfer rate.
\item We find that the light curves show enigmatic obscuration events consisting of rapid drops of brightness of up to $\Delta$$V$ = 0.3 mag. These events might be related to inhomogeneities present in the flow of mass loss but more studies are needed to clarify their nature.

\end{itemize}

\section*{Acknowledgements}
 We thank the anonymous referee that helped to improve the first version of this manuscript.
R. E. M.,  J. G. and M. C. acknowledge support by  BASAL Centro de Astrof{\'{i}}sica y Tecnolog{\'{i}}as Afines (CATA) PFB-06/2007 and FONDECYT 1190621. JG acknowledges ANID project 21202285. 
G. D., I. V., J. P. and M. I. J.  acknowledge the financial support of the Ministry of Education, Science and Technological Development of the Republic of Serbia through the contract No 451-03-68/2020/14/20002. 
M. C. and P. H. acknowledge support by  Astronomical Institute of the Czech Academy
of Sciences through the project RVO 67985815.
D. K. thanks GA 17-00871S of the Czech Science Foundation. This research has made use of: 
(1) the SIMBAD database, operated at CDS, Strasbourg, France 
\citep{2000A&AS..143....9W}, (2) data products from the Wide-field Infrared
Survey Explorer, which is a joint project of the University of California, Los
Angeles, and the Jet Propulsion Laboratory/California Institute of Technology,
funded by the National Aeronautics and Space Administration, (3)
the NASA/ IPAC Infrared Science Archive, which
is operated by the Jet Propulsion Laboratory, California Institute of
Technology, under contract with the National Aeronautics and Space
Administration, (4) data products from the Two Micron All Sky Survey, which is a joint project of the University of Massachusetts and the Infrared Processing and Analysis Center/California Institute of Technology, funded by the National Aeronautics and Space Administration and the National Science Foundation and (5) data from the OMC Archive at CAB (INTA-CSIC), pre-processed by ISDC.








\label{lastpage}

\end{document}